# Real-time Electrical Tuning of an Optical Spring on a Monolithically Integrated Ultrahigh Q Lithium Nibote Microresonator

Zhiwei Fang,[1,2,†] Sanaul Haque,[3,†] Jintian Lin,[4,†] Rongbo Wu,[4] Jianhao Zhang,[4] Min Wang,[1,2] Junxia Zhou,[1,2] Muniyat Rafa,[3] Tao Lu,[3,5] Ya Cheng,[1,2,4,6]

[1]*State Key Laboratory of Precision Spectroscopy, East China Normal University, Shanghai 200062, China.*
[2]*XXL—The Extreme Optoelectromechanics Laboratory, School of Physics and Materials Science, East China Normal University, Shanghai 200241, China.*
[3]*Department of Electrical and Computer Engineering, University of Victoria, Victoria, BC, V8P 5C2, Canada.*
[4]*State Key Laboratory of High Field Laser Physics, Shanghai Institute of Optics and Fine Mechanics, Chinese Academy of Sciences, Shanghai 201800, China.*
[5]*taolu@ece.uvic.ca*
[6]*ya.cheng@siom.ac.cn*

**Cavity optomechanics, the study of the interplay between light and mechanical properties of matter, has triggered a wide range of groundbreaking researches from cavity quantum electrodynamics, label free single molecule detection to the creation of phonon laser. Using femtosecond laser direct writing followed by chemo-mechanical polishing, here we report an ultrahigh quality (Q~$10^7$) factor lithium niobate (LN) whispering gallery microresonator monolithically integrated with in-plane microelectrodes. Coherent regenerative optomechanical oscillation with an effective mechanical quality ($Q_m$) factor as high as $2.86 \times 10^8$ is observed in air. We demonstrate real-time electrical tuning of the optomechanical frequency with an electro-mechanical tuning efficiency around -134 kHz/100V.**

## 1. INTRODUCTION

The whispering gallery mode (WGM) optical microresonator, due to its capacity to forming an optical spring driven by the enormous optical force at resonance, has become an excellent platform for studying optomechanics [1]. Optical microresonators of various materials such as silica, silicon, silicon carbide, silicon nitride, aluminum nitride, gallium arsenide, gallium phosphide and diamond have been demonstrated as chip-scale optomechanical systems and applied to sensing, information processing and quantum physics [2-14]. Compared with these materials, lithium niobate (LN) is advantageous for its broad optical transparency window (0.35–5 μm), high nonlinear coefficient ($d_{33} = -41.7 \pm 7.8 \ pm/V@\lambda = 1.058 \ \mu m$), high refractive index (~2.2) and large electro-optical effect ($r_{33} = 30.9 \ pm/V@\lambda = 632.8 nm$) [15]. These outstanding physical properties collectively enhance the optomechanical responses in LN WGM microresonator, and in turn provide opportunities to new findings and applications. Particularly, the advent of LN on insulator (LNOI) and the development of micro/nano-fabrication techniques enables fabrication of on-chip LN microresonators with high optical quality factors (Q), thereby dramatically enhancing the strength of optical field in the cavity. However, until recently the fabrication of LNOI nanophotonic structures exclusively relies on ion etching, as the crystalline LN cannot be easily patterned with the traditional optical lithography owing to its high chemical stability [16-24]. The ion etching inevitably leaves a surface roughness on the order of a few nanometers, limiting the optical Q factor of a freestanding LN microdisk to ~$10^6$ [18-22]. The problem has recently been resolved by first patterning a chromium (Cr) thin film on the top surface of LNOI with a femtosecond laser, followed by the chemo-mechanical polishing (CMP) for structuring the LNOI into the microdisk [25]. The CPM gives rise to extremely high surface quality with a roughness as low as 0.452 nm [26], allowing for promoting the Q factor of the LN microresonator by one order of magnitude as compared to the ionically etched LN microresonators.

In this article, we report an ultrahigh quality factor LN microresonator monolithically integrated with in-plane microelectrodes. Due to the ultrahigh optical Q factor exceeding 10 million, even in the highly viscous air environment, coherent regenerative optomechanical oscillation occurs at a threshold optical power as low as 1.7 mW. We further record an ultrahigh effective mechanical quality ($Q_m$) factor of $2.86 \times 10^8$ as compared with the previous results [23,27]. More importantly, we demonstrate for the first time the real-time electrical tuning of the optomechanical frequency with an electro-mechanical tuning efficiency around -134 kHz/100V.

## 2. DEVICE FABRICATION

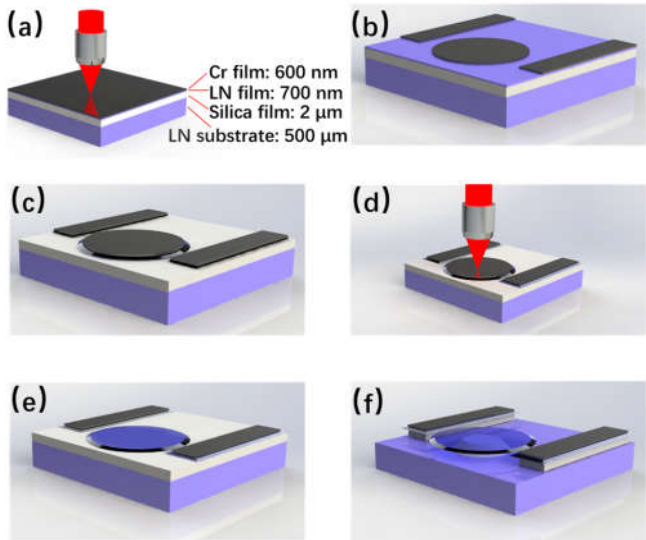

Fig. 1. The processing flow of fabricating an on-chip LN microresonator integrated with electrodes is illustrated: (a)-(b) Patterning the Cr thin film into the circular disk and rectangular electrodes using femtosecond laser microfabrication. (c) Transferring the circular (microresonator) and rectangular (electrode) shaped patterns to LNOI by chemo-mechanical polishing. (d) Selective removal of Cr film on the LN mircoresonator using femtosecond laser ablation. (e)-(f) Chemical wet etching of the sample to form the freestanding LN microresonator.

In our experiment, the on-chip LN microresonator integrated with Cr electrodes was fabricated on a commercially available Z-cut LN thin film wafer with a thickness of 700 nm (NANOLN, Jinan Jingzheng Electronics Co., Ltd). The LN thin film is bonded by a silica layer with a thickness of ~2 μm on a 500 μm thick LN substrate, and a 600-nm-thickness layer of chromium (Cr) film was deposited on the surface of the LNOI by magnetron sputtering method.

The fabrication process includes four steps, as illustrated in Fig. 1. First, the Cr film on the LNOI sample was patterned into a circular resonator and two rectangular electrodes using space-selective femtosecond laser (laser source: Pharos-6W, center wavelength: 1026 nm, pulse width: 170 fs) direct writing. Specifically, to minimize the heat effect as well as the redeposition of the ablation debris on the disk surface, the femtosecond laser ablation was conducted by immersing the sample in water. An objective lens (100×/NA 0.7) was used to focus the femtosecond laser beam down to a ~1 μm-diameter focal spot. The sample could be arbitrarily translated in 3D space at a resolution of 100 nm using a PC-controlled XYZ stage combined with a nano-positioning stage (ABL15020, Aerotech Inc.,). More details of the femtosecond laser micromachining can be found in Ref. 25.

Subsequently, the chemo-mechanical polishing (CMP) process was performed to fabricate the LN microresonator and Cr electrodes by a wafer polishing machine (NUIPOL802, Kejing, Inc.). In the CMP process, a piece of velvet polishing cloth and the colloidal silica polishing suspension (MasterMet, Buehler, Ltd) were used. Next, the femtosecond laser microfabrication was performed again to selectively remove the Cr film on the LN mircoresonator. Finally, the fabricated structure was immersed in a buffered hydrofluoric acid (HF) solution (BUFFER HF IMPROVED, Transene Co., Inc.) for 20 mins to partially under etch the $SiO_2$ layer into the shape of a pillar supporting the LN microdisk. It takes about 2 hrs in total to produce the whole integrated device.

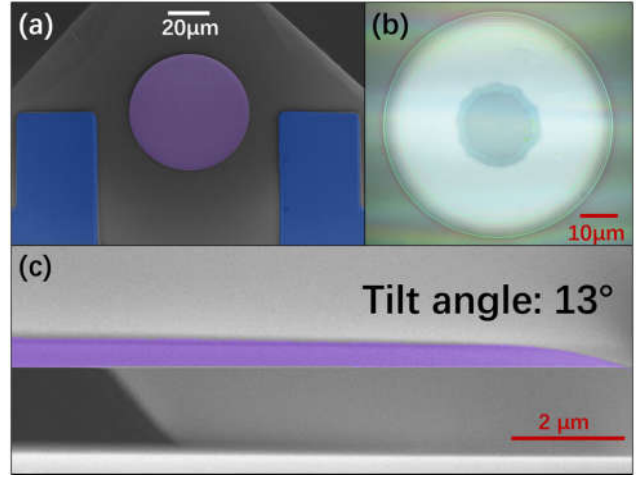

Fig. 2. (a) Top view false color SEM of the 66-μm LN microresonator (purple) integrated with Cr electrodes (blue) and (b) optical micrograph of the LN disk. (c) Side view SEM of the fabricated LN microresonator with a wedge angle of 13°.

Figure 2(a) presents the false color scanning electron micrograph (SEM) where the LN microresonator is shown in purple and the microelectrodes are shown in blue. The top view optical micrograph in Fig. 2(b) shows the freestanding LN microdisk with a diameter of 66 μm supported by a fused silica pillar with a diameter of ~20 μm. The rim of the LN microdisk displays interference patterns under illumination, indicating the varying thickness at the edge of the LN disk. The wedge shape of the disk edge is further confirmed in Fig. 2(c) where the wedge angle of 13° is observed. The extended wedge helps to enhance the optical Q factors as the light scattering from the edge tip is significantly reduced [28]. The two Cr microelectrodes in the rectangular shape are separated by 15 μm from the LN microresonator. It is worth mentioning that the microelectrodes are patterned by the femtosecond laser micromachining with the LN microdisk at the same fabrication step. This gives rise to a high fabrication efficiency and a high fabrication resolution. The small distance between the microelectrodes enabled by the high precision femtosecond laser micromachining is critical for achieving the highly efficient electro-optical tuning of both the optical and mechanical modes.

## 3. DEVICE CHARACTERIZATION

To characterize the optical and mechanical modes of the integrated LN optomechanical system, we used a setup as shown in Fig. 3(a). Here, a tunable laser (TLB, New Focus Inc.) was used to couple the light into and out of the fabricated LN microresonator through a tapered fiber with a waist of 1 μm. The polarization states were selectively excited using an in-line fiber polarization controller. A photoreceiver (New focus 1811-FS) was connected to the output of the tapered fiber for capturing the optical signal. The captured signal is then converted to electrical signals and analyzed by an oscilloscope (Tektronix MDO3104) and a real-time spectrum analyzer (Tektronix RSA5126B). An open-loop piezo controller (Core tomorrow Co. Ltd, Model A series) was used as the voltage supply for Cr electrodes, which provided a variable voltage ranging from 0 V to 900 V. In particular, from the inset in Fig. 3(a), the 3D distribution of electric field in the LN microresonator was simulated by using a finite element method (COMSOL Multiphysics). The calculated capacitance between the two Cr microelectrodes is about $C=1.5668\times10^{-16}$ F.

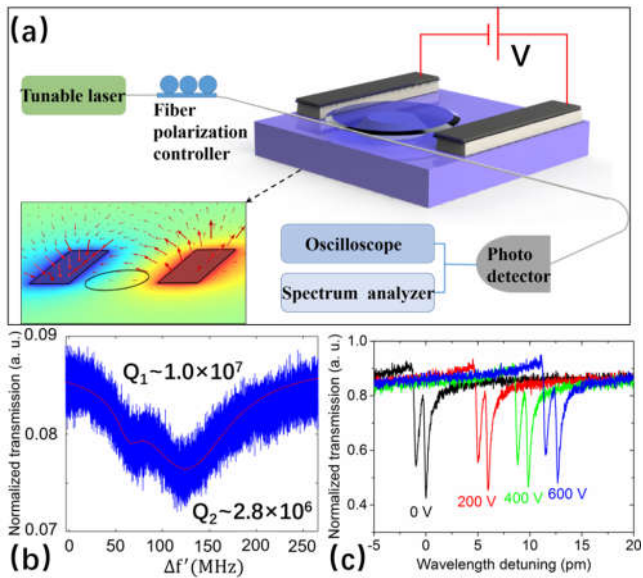

Fig. 3. (a) Experimental setup for characterizing the transmission spectra and mechanical mode of the optomechanical system. Inset (lower left): simulated 3D distribution of the electric field in LN microresonator. (b) The double Lorentzian fitting showing a mode splitting, indicating the Q-factors of $1.0\times10^7$ and $2.8\times10^6$ as measured around 970 nm wavelength. (c) The resonant wavelength shifts linearly with an electro-optical tuning efficiency of ~2.5 pm/100V.

As shown in Fig. 3(b), a mode splitting was observed. Based on a double Lorentzian fitting, the Q factors of the splitting modes were determined to be $1.0\times10^7$ and $2.8\times10^6$ around 970 nm wavelength. Figure 3(c) shows that the resonant wavelength continuously shifts to the red side of the spectrum with the increase of the applied direct current (DC) voltage, and the electro-optical tuning efficiency is determined to be ~2.5 pm/100V.

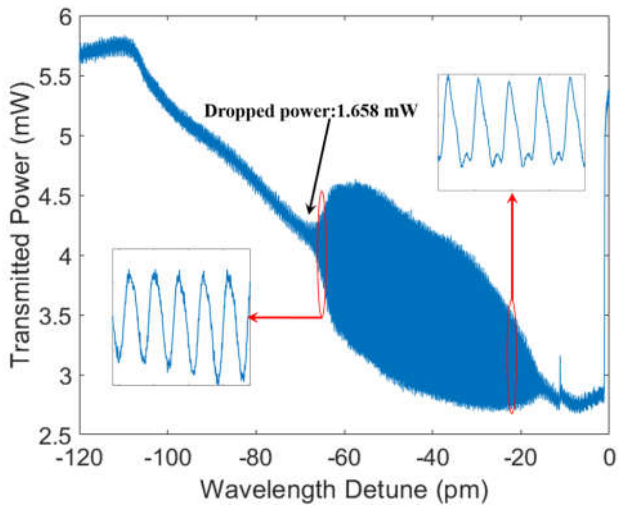

Fig. 4. Transmitted optical power as a function of probe laser wavelength detune. At a dropped optical power close to the threshold power, the oscillation was observed. The bottom left inset of the plot displayed a sinusoidal spectrum; whereas at a high dropped power, the spectrum displayed in the top right inset of the plot was distorted by the high order harmonics.

During the optomechanical experiment, we linearly scanned the optical frequency of the probe laser using a waveform generator and adjusted the coupling between the tapered fiber and the LN microresonator such that the oscillation was observable as shown in Fig. 4. When a power drop in the cavity was observed around ~1.7 mW, regenerative coherent optomechanical oscillation was observed and details of the transmission spectra at different dropped powers were displayed in the two insets.

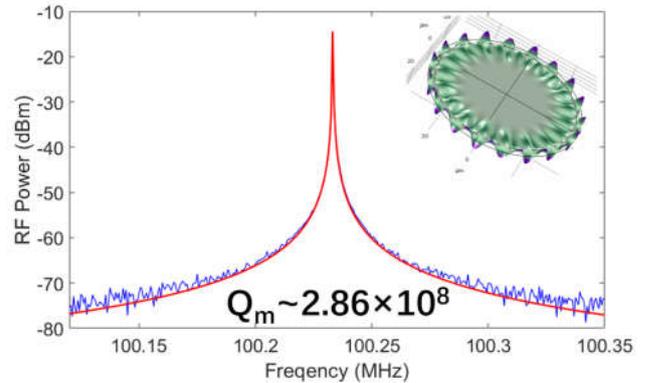

Fig. 5. Radio Frequency (RF) spectrum of the cavity transmission showing the oscillation frequency at 100.23 MHz, with the theoretical fitting in red. The top right inset is the simulation result of the mechanical mode in the LN disk, of which the oscillation frequency is calculated to be 100.79 MHz, in good agreement with the measurement.

By switching the laser into continuous wave mode (CW) and tuning the wavelength into the cavity resonance, a mechanical spectrum captured by the spectrum analyzer shows an effective mechanical $Q_m$~$2.86\times10^8$ at the mechanical frequency of 100.23 MHz at a low dropped optical power of 2.94 mW, corresponding to an oscillation linewidth as narrow as 0.35 Hz. The effective mechanical quality factor in our experiment is significantly higher than that reported in a LN based optomechanical system [23,27]. The mechanical mode can be visualized with the simulation as shown in the inset of Fig. 5. The mechanical frequency of 100.79 MHz obtained in the simulation agrees well with the experimental measurement.

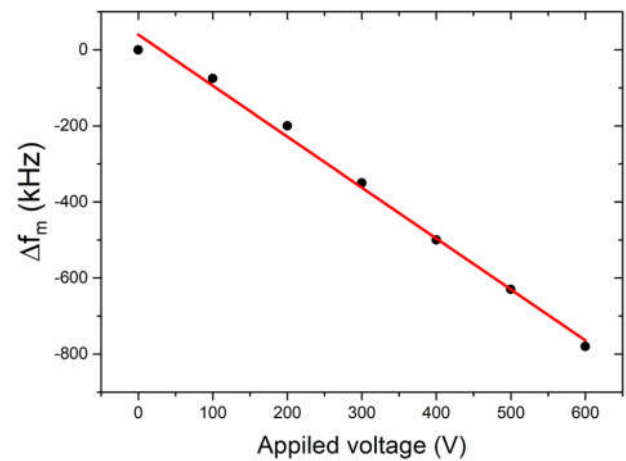

Fig. 6. Mechanical frequency decreases linearly by 75 kHz, 200 kHz, 350 kHz, 500 kHz, 630 kHz, and 780 kHz at the DC voltages of 100 V, 200 V, 300 V, 400 V, 500 V, and 600 V, respectively.

To demonstrate the real-time electrical tuning of the mechanical mode, we recorded the mechanical frequency at the different electric voltages. Figure 6 shows that the measured mechanical frequency decreases monotonically with the increase of the voltage applied to the microelectrodes. The mechanical frequency decreases by 75 kHz, 200 kHz, 350 kHz, 500 kHz, 630 kHz, and 780 kHz, at the voltages of 100V, 200V, 300V, 400V, 500V, and 600V respectively. The linear fitting of the measured data indicates an electro-mechanical tuning efficiency ($\frac{d\Delta f}{dV}$) of ~-134 kHz/100V.

## 4. CONCLUSION

To conclude, we demonstrate an electrically tunable monolithically integrated optomechanical system which possesses simultaneously ultrahigh optical Q and ultrahigh mechanical Q at high mechanical frequencies. Our result suggests that LNOI is an ideal candidate for optomechanic system. This is particularly true when high optical Q LNOI disks are now readily achievable with the CMP-based lithography. The demonstrated optomechanical device enables great potential for a broad range of applications in metrology, sensing, information processing and quantum physics.

**Funding**. National Natural Science Foundation of China (Grant Nos. 11734009, 11874375, 61590934, 61635009, 61327902, 61505231, 11604351, 11674340, 61575211, 61675220, 61761136006), the Strategic Priority Research Program of Chinese Academy of Sciences (Grant No. XDB16000000), the Key Research Program of Frontier Sciences, Chinese Academy of Sciences (Grant No. QYZDJ-SSW-SLH010), the Project of Shanghai Committee of Science and Technology (Grant No. 17JC1400400), the Shanghai Rising-Star Program (Grant No. 17QA1404600), and the Shanghai Pujiang Program (Grant No.18PJ1403300). NSERC Discovery (Grant No. RGPIN-2015-06515). We would like to acknowledge CMC Microsystems and CAMTEC.

† These authors contributed equally to this work.